# Ultrafast structural color change in indium tin oxide/titanium dioxide 1D photonic crystal

Liliana Moscardi [1,2], Stefano Varas [3], Alessandro Chiasera [3], Francesco Scotognella [1,2], Michele Guizzardi [1]

[1] Dipartimento di Fisica, Politecnico di Milano, Piazza Leonardo da Vinci 32, 20133 Milano, Italy
[2] Center for Nano Science and Technology@PoliMi, Istituto Italiano di Tecnologia (IIT), Via Giovanni Pascoli, 70/3, 20133, Milan, Italy
[3] Istituto di Fotonica e Nanotecnologie IFN – CNR, Via alla Cascata, 56/C, 3812, Povo – Trento, Italy

**Abstract**
Photonic crystals can integrate plasmonic materials such as indium tin oxide (ITO) in their structure. Exploiting ITO plasmonic properties it is possible to tune the photonic band gap of the photonic crystal upon the application of an external stimuli. In this work, we have fabricated a one-dimensional multilayer photonic crystal via radiofrequency sputtering and we have triggered its optical response with ultrafast pump-probe spectroscopy. Upon photoexcitation we observe a change in the refractive index of indium tin oxide. Such effect has been used to create a photonic crystal that change its photonic bandgap in an ultrafast time scale. All optical modulation in the visible region, that can be tuned by designing the photonic crystal, has been demonstrated.

**Introduction**:
Photonic crystals represent a class of periodic structures that can be integrated into a wide range of technologies. Thanks to their high symmetry (in one, two or three dimensions) and the difference in terms of refractive indices of the constituent materials, an interference of the incident and diffused electromagnetic wave is generated, causing the formation of a forbidden band for the passage of photons, called photonic band gap) [1–4]. This attributes to photonic crystals their peculiar optical properties and the generation of their changing colors, called structural colors. There are several applications in which multilayer photonic crystals can be exploited, such as distributed feedback lasers [5–8], perfect absorbers [9,10], and sensors [11–15]. Selecting responsive materials for their fabrication, it is possible to modulate their optical response upon application of external stimuli. Photonic crystals can be fabricated with several techniques in one dimension, two dimensions and three dimensions [16]. Moreover, quasicrystals [17] and disordered structures [18,19] can be fabricated.
There are several applications in which multilayer photonic crystals can be exploited, such as distributed feedback lasers [9–12], perfect absorbers [13,14], and sensors [15–19]. Multilayer photonic crystals, also called Distributed Bragg Reflectors (DBR),that include indium tin oxide are promising devices since this plasmonic material is able to show a high refractive index change with an applied electric field [20]. ITO-based DBRs take advantage of the ability to change their refractive index through the application of an electric field in order to modulate the position of the PBG. The optical properties of the multilayer can be easily tuned by acting on the refractive index of the constituent materials or on the geometric parameters of the lattice. [21,22].
In this work we have fabricated a one dimensional photonic crystal (DBR) made by indium tin oxide and titanium dioxide via radiofrequency sputtering. The fabricated multilayer photonic crystal shows a very high optical quality with a strong photonic band gap. By pumping the photonic crystal at 1550 nm, in the range of the plasma frequency of ITO, we observe in transient differential transmission measurements recombination dynamics within the first picosecond of time delay. Such dynamics is mainly due to electron-phonon scattering in ITO. The ultrafast modulation of the plasma frequency of ITO results in an ultrafast modulation of the photonic band gap of the ITO-based photonic crystal.

**Experimental methods**

*Sample preparation*: The multilayer ITO/TiO$_2$ structures were deposited on silicon, vitreous silica glass (v-SiO$_2$) by RF magnetron sputtering. The v-SiO$_2$ substrates have dimensions 75 mm × 25 mm × 1 mm. Before deposition, the substrates were ultrasonically cleaned in deionized water then they were cleaned with ethanol and finally dried in nitrogen. The substrates were after cleaned inside of the RF sputtering chamber for 10 min just before starting the deposition while heating up the temperature to 120 °C at a pressure of 10$^{-6}$ mbar.

The magnetron RF sputtering deposition of ITO and TiO$_2$ films was performed by alternatively changing a 15 cm × 5 cm ITO and 15 cm × 5 cm TiO$_2$ targets. The residual pressure before the deposition was 4.5×10$^{-7}$ mbar. During the deposition procedure, the substrates were not heated, and the sample holder temperature was kept at 30 °C. The sputtering was performed in an Ar atmosphere (5.4×10$^{-3}$ mbar) and an RF power of 110 W was applied on TiO$_2$ target and 80 W applied on the ITO target. To monitor the thickness of the layers during the deposition, two quartz microbalances INFICON model SQM-160, faced on the two targets were employed. The final resolution on the effective thickness obtained by these quartz microbalances is about 1 Å. More details are available in reference [19]. The deposited structure consists of 5 couple ITO/TiO$_2$ for a total of 10 layers. The first layer deposited directly on the substrates is ITO while the last layer is TiO$_2$. Reference ITO and TIO$_2$ single layer were also fabricated using the same deposition protocol on silicon and vitreous silica glass substrates.

*Optical characterization*: The steady-state light transmission spectrum of the photonic crystal has been acquired with a JASCO spectrophotometer. The ultrafast differential transmission measurements have been performed by using a Coherent Libra amplified laser system with the fundamental wavelength at 800 nm, a pulse duration of about 100 fs and a repetition rate of 1 kHz. Noncollinear optical parametric amplifiers to tune the pump wavelength has been built with a procedure reported in Reference [23]. White light generation for the probe pulse has been achieved focusing the fundamental beam into a sapphire plate. The differential transmission $\Delta T/T = (T_{ON} - T_{OFF})/T_{OFF}$ has been acquired with an optical multichannel analyzer. $T_{ON}$ and $T_{OFF}$ indicate the probe spectra transmitted the excited and unperturbed sample, respectively.

**Results and Discussion**

In Figure 1 we can see the transmission of the ITO layer, which is transparent in the visible and we see decrease of transmittivity below 400 nm due to intragap state [24] and the more evident decrease where we have the band gap transition around 340 nm. Titanium dioxide is completely transparent in the visible and have a bandgap around 330 nm [25] so is completely covered by glass adsorption that can be seen in Figure 1. As we discussed above the staking of two dielectric materials (ITO and TiO$_2$) creates a Photonic Band Gap that completely changes the transmission compared to the one of the single layers, we can see the appearance of some dip where the transmission of the photon inside the structure is neglected.

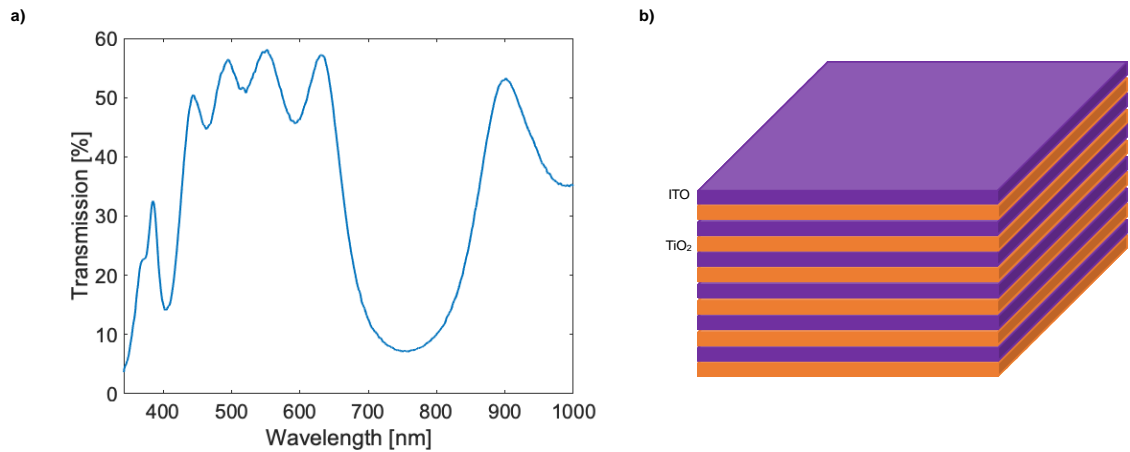

*Figure 1 a) Steady-state transmission spectrum of the ITO/TiO$_2$ PC. b) Sketch of the one-dimensional multilayer photonic crystal.*

We want to study how the transmission through the PC change upon photoexcitation of the plasmon. To do so we tune our OPA at 1550 nm where we have the ITO plasmonic absorption.
After the excitation of the plasmon in a timescale of about 10 fs due to electron scattering we have the plasmon dephasing and the creation of a non-thermalized Fermi distribution (FD). After this, within 100 fs a hot FD is be created by electron-electron scattering. With our time resolution of about 100 fs we cannot see those events, but we can follow the subsequent relaxation of the hot-FD through electron-phonon scattering, that, happens in a timescale of less than 1 ps. After this a much slower process take place which is the phonon-phonon scattering to get rid of the heat of the lattice.

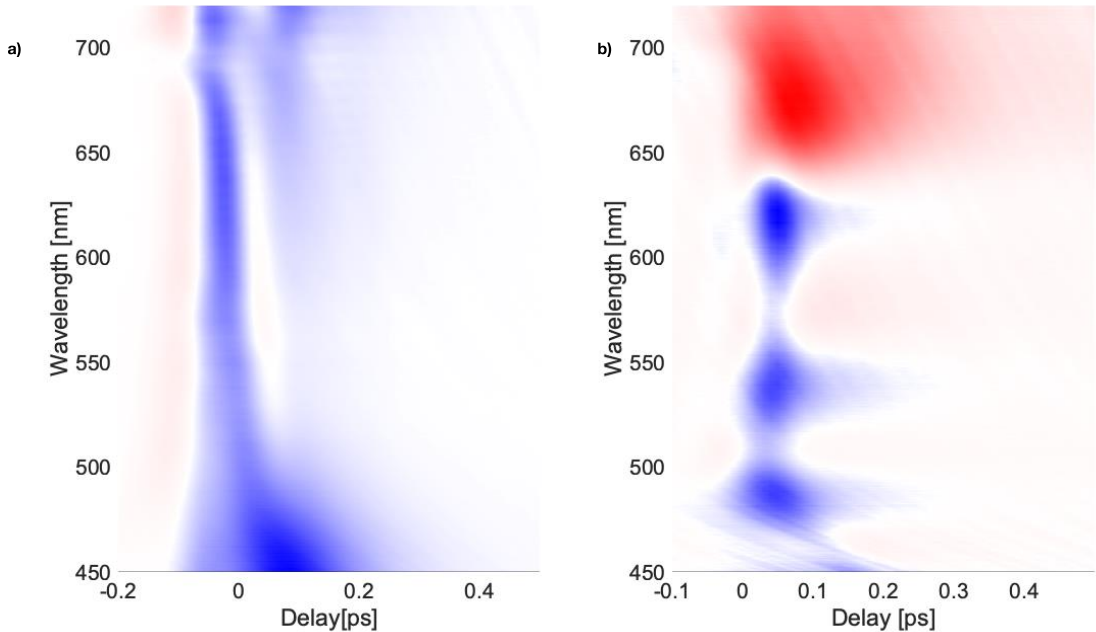

*Figure 2 a) Transient response of ITO. b) Transient response of the PC structure ($\Delta T/T > 0$ in red, $\Delta T/T < 0$ in blue).*

In Figure 2a we can see the transient response of the ITO layer. Around zero time delay we have the so-called Cross phase modulation artifact (XPM), that is responsible for the initial positive-negative-positive signal. As we discussed above the recombination dynamics are faster than 1 ps and can be seen in Figure 2a. We see a negative response in the whole visible spectra because by exciting the plasmon and creating the hot FD we are changing the dielectric response of the material, by changing both the plasmon resonance and the refractive index [26,24]. This leads to an ultrafast change in the reflectivity of the ITO layer, and this is seen as a negative signal in terms of $\Delta T/T$ because more photons are reflected at the air-ITO interface.

This negative signal increase toward shorter wavelength, when we are around 440 nm we have a strong negative signal induced by the so-called inverse Moss-Burstein effect: before excitation all the transition from the intragap state and the conduction band are neglected because the levels are already filled with electrons, while when we create the hot-FD we free some space and those transition are now allowed resulting in a less transmitted probe.

In Figure 2b we can see the response of the photonic crystal, now we can see a completely different response that has positive and negative features. As we discussed above, this effect is induced by the change in the refractive index of the ITO. This modulation is still faster than 1 ps because it arises from the plasmon decay.

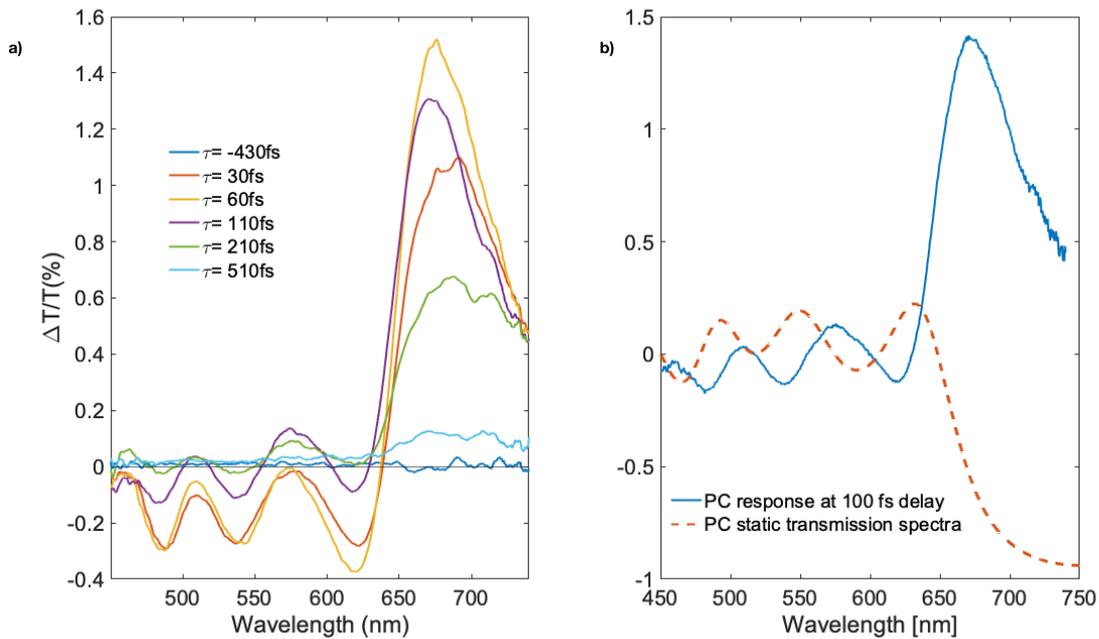

*Figure 3 a) Spectra at fixed time delay of the PC. b) Static spectra and differential transmittivity signal of the structure.*

In Figure 3a we can better appreciate the evolution of the spectral response at some fixed delay time, at around 510 fs the modulation is basically disappeared. In Figure 3d we depicted the modulation at 100 fs delay and the static transmission spectra of the PC. We can see that we have an ultrafast red shift of the photonic band gap.
By changing the thickness, refractive index and the number of layers of the PC we can change the complexity and position of the resonance in the photonic bandgap. With this system we can modulate a visible beam by using an infrared pulse and the complexity of the modulation can be tuned accordingly. The de-excitation of the plasmon is faster than 1 ps, this gives us the capability to modulate an incoming visible beam with a repetition rate faster than 1 THz.

**Conclusion**
The differential transmission measurements show recombination dynamics faster than 1 ps. The change in the refractive index of ITO upon photoexcitation of the plasmon resonance can used to create a photonic crystal that change its photonic bandgap in an ultrafast time scale. This allow us to achieve all optical modulation in the visible region that can be tuned by designing the physical parameter of the PC accordingly.

**Acknowledgement**
This project has received funding from the European Research Council (ERC) under the European Union's Horizon 2020 research and innovation programme (grant agreement No. [816313]).